# NUMERICAL INVESTIGATION OF EVAPORATION INDUCED SELF-ASSEMBLY OF SUB-MICRON PARTICLES SUSPENDED IN WATER

**Raihan Tayeb, Yijin Mao, and Yuwen Zhang[1]**
Department of Mechanical and Aerospace Engineering
University of Missouri
Columbia, MO 65211, USA

**ABSTRACT**

Self-assembly of sub-micron particles suspended in a water film is investigated numerically. The liquid medium is allowed to evaporate leaving only the sub-micron particles. A coupled CFD-DEM approach is used for the simulation of fluid-particle interaction. Momentum exchange and heat transfer between particles and fluid and among particles are considered. A history dependent contact model is used to compute the contact force among sub-micron particles. Simulation is done using the open source software package CFDEM which basically comprises of two other open source packages OpenFOAM and LIGGGHTS. OpenFOAM is a widely used solver for CFD related problems. LIGGGHTS, a modification of LAMMPS, is used for DEM simulation of granular materials. The final packing structure of the sub-micron particles is discussed in terms of distribution of coordination number and radial distribution function (RDF). The final packing structure shows that particles form clusters and exhibit a definite pattern as water evaporates away.

## INTRODUCTION

Self-assembly of colloids and micro-scale particles has become a subject of great interest due to widespread advancement of micro-scale technologies. Fabrication of micro-scale structures essential for many industrial applications such as optics [1-3], catalysis [4, 5] and printing technology [6] can be achieved by assembling micro-sized particles. Self-assembly of micro-scale particles induced by evaporation is a simple yet widely used microfabrication technique [7-10]. A complete numerical analysis of such a phenomena can be of great importance by providing deeper understanding of the assembling process. Many numerical methods such as Lattice Boltzmann method (LBM) [11, 12], Immersed Moving Boundary (IMB) method [13] are used to simulate the interaction between particle and fluid. However these methods have the disadvantage that if the number of particle is large computational efficiency increases enormously. The method used in this paper CFD-DEM is a robust and efficient method for simulating interactions among fluid and large number of particles. In CFD-DEM, particle to particle interaction is computed by Discrete Element method (DEM) which is a highly appreciated Lagrangian approach for particle interaction simulation. The fluid phase is governed by Navier Stokes equation and finite volume method is used for the CFD part.

In this study the final packing structure of 4500 mono-sized particles with radius of 5 μm is analyzed after they are settled down and the liquid medium in which the particles are suspended is evaporated. While most researchers focus on the momentum exchange between particle and fluid [14-16], in this study the heat transfer between particles and fluid and among particles are also taken into account. The problem in this case gets more challenging as evaporation or phase change is also included. The final packing structure of the particles is analyzed using distribution of coordination number and radial distribution function (RDF). Just before the liquid evaporates the particles are seen to agglomerate and form a pattern.

## NOMENCLATURE

| | |
|---|---|
| $A_s$ | spherical area, m$^2$ |
| $c$ | damping coefficient, s |
| $c_p$ | specific hea, J/Kg K |
| $C$ | Cunningham correction factor |
| $C_e$ | evaporation coefficient |
| $d$ | diameter of particle, m |
| $D$ | distance, m |
| $e$ | coefficient of restitution |
| $E$ | elastic modulus, Pa |
| **F** | force on particle, N |
| **g** | gravity, m/s$^2$ |
| $h$ | heat transfer coefficient, W/m$^2$ K |
| $h_e$ | enthalpy of evaporation, J/kg |
| $I$ | moment of inertia, kg·m$^2$ |
| $k_f$ | thermal conductivity of fluid, W/m K |
| $k_s$ | thermal conductivity of particle, W/m K |
| $m$ | mass of particle, kg |
| $M$ | molecular mass of water, kg |
| **n** | unit vector, m |
| $Nu$ | Nusselt number |
| $p$ | pressure, Pa |
| Pr | Prandtl number |
| $Q$ | power, W |
| **r** | position vector, m |

---

| **R** | position vector, m |
|---|---|
| $R$ | radius of particle, m |
| $R$ | universal gas constant, J/mol K |
| Re | Reynold number |
| $S$ | source term |
| $t$ | time, s |
| **T** | torque, N m; temperature, K |
| $T_{sat}$ | saturation temperature, K |
| **U** | velocity of fluid, m/s |
| $\mathbf{U_r}$ | relative velocity, m/s |
| $u_p$ | velocity of particle, m/s |
| $V_s$ | volume of particle, m$^3$ |
| $V_c$ | volume of CFD cell, m$^3$ |

**Greek Symbols**

| $\alpha_f$ | volumetric fraction of fluid |
|---|---|
| $\gamma$ | surface energy density, J/m$^2$ |
| $\mu$ | fluid viscosity, Pa s |
| $\delta_n$ | normal direction displacement, m |
| $\delta_t$ | tangential displacement, m |
| $\lambda$ | mean free path of water, m |
| $\mu_s$ | sliding friction coefficient |
| $\mu_r$ | rolling friction coefficient |
| $\rho$ | fluid density, kg/m$^3$ |
| $\sigma$ | Poisson ratio; surface tension, N/m |
| $\tau$ | shear stress tensor, Pa |
| $\omega$ | angular velocity, rad/s |

## METHODS AND PHYSICAL MODELS

The governing equation for the fluid phases consists of a set of volume averaged mass and momentum balance equations in an Eulerian description. The fluid phases are considered incompressible and in the CFD domain the particle phase is characterized by void fraction field $\alpha_f$. The liquid and the vapor phases are denoted by $\alpha_1$ and $\alpha_2$ respectively.

The mass and momentum equations are given as:

For fluid phase:

$$\frac{\partial(\rho\alpha_f)}{\partial t} + \nabla \cdot (\rho\alpha_f \mathbf{U}) = 0 \tag{1}$$

$$\frac{\partial(\alpha_1 \alpha_f)}{\partial t} + \nabla \cdot (\alpha_1 \alpha_f \mathbf{U}) + \nabla \cdot (\alpha_1 \alpha_f \mathbf{U}_r (1-\alpha_1)) = \dot{S}_v \tag{2}$$

$$\frac{\partial}{\partial t}(\alpha_f \rho \mathbf{U}) + \nabla \cdot (\alpha_f \rho \mathbf{U} \otimes \mathbf{U}) = -\nabla \cdot (\alpha_f p_{rgh}) - \mathbf{g} \cdot \mathbf{r} \nabla(\alpha_f \rho) + \nabla \cdot (\alpha_f \tau) + \alpha_f \sigma_K \nabla \alpha_1 - \mathbf{F}_d \tag{3}$$

where

$$\alpha_1 \alpha_f + \alpha_2 \alpha_f = \alpha_f \tag{4}$$

$$\alpha_1 + \alpha_2 = 1 \tag{5}$$

$$\alpha_f p_{rgh} = \alpha_f p - \alpha_f \rho \mathbf{g} \cdot \mathbf{r} \tag{6}$$

$$\tau = \mu\left(\nabla \mathbf{U} + (\nabla \mathbf{U})^T\right) - \frac{2}{3}\left[\mu(\nabla \cdot \mathbf{U})I\right] \tag{7}$$

where $\rho$ is the average fluid density, **U** is the average velocity of fluid, $\dot{S}_v$ represents source terms due to phase change, $\sigma_K$ is the surface tension effect at the interface, $p_{rgh}$ is the dynamic pressure, $\mathbf{F}_d$ is the drag force term, $\mu$ is the mean viscosity and **g** is the acceleration due to the gravity. The third term on the left side of equation (2) is the compressibility term added to obtain sharp interfaces [17] and $\mathbf{U}_r$ is the relative velocity between two fluid phases.

For solid phase:

$$m_i \dot{\mathbf{u}}_{p,i} = \mathbf{F}_{ij}^n + \mathbf{F}_{ij}^t + \mathbf{F}_{coh} + m_i \mathbf{g} + \mathbf{F}_d + \mathbf{F}_b \tag{8}$$

$$I_i \dot{\boldsymbol{\omega}}_i = \mathbf{T}_{ij}^r + \mathbf{T}_{ij}^t \tag{9}$$

where $m_i$ is the mass of the i$^{th}$ particle, $\mathbf{u}_{p,i}$ is the velocity of the i$^{th}$ particle, $I_i$ is the moment of inertia of the particle and $\boldsymbol{\omega}_i$ is the angular velocity of the i$^{th}$ particle. $\mathbf{F}_{ij}^n$ and $\mathbf{F}_{ij}^t$ are normal and tangential force due to contact, $\mathbf{F}_{coh}$ is the cohesive force between two particles, $m_i \mathbf{g}$ is force of gravity, $\mathbf{F}_d$ is the drag force due to relative fluid motion and $\mathbf{F}_b$ is the buoyant force. $\mathbf{T}_{ij}^r$ is the torque due to rolling friction and $\mathbf{T}_{ij}^t$ is the torque due to tangential force. The models used to depict the forces will be introduced as follow:

Drag force model:

The drag force accounts for the force acting on the particle due to relative motion between particle and fluid. The drag model used is suitable for particles in the near sub-micron range [18].

$$\mathbf{F}_d = \frac{18\mu}{d^2 C_c}(\mathbf{U} - \mathbf{u}_p) \tag{10}$$

where

$$C_c = 1 + \frac{2\lambda}{d}\left(1.257 + 0.4 e^{-(1.1d/2\lambda)}\right) \tag{11}$$

where $\lambda$ is the mean free path of water which is $2.5 \times 10^{-10}$ m and $d$ is the diameter of the particle.

Buoyant force:

$$\mathbf{F}_b = -\mathbf{g}\rho V_s \tag{12}$$

where $V_s$ is the volume of the particle.

Contact force model:

A contact force model is implemented to take into account the interaction among particles and between particle and wall. The model used is the Gran-Hertz-History model where a non-linear relationship is assumed between overlap distance during collision and normal contact force. Other force that are included are viscous damping force and cohesive force. Tangential force during collision is also taken into account.

The normal contact force during collision $\mathbf{F}_{ij}^n$ is given by [19-22]:

$$\mathbf{F}_{ij}^n = \left[Y_n \delta_n - c_n (\mathbf{v}_{ij} \cdot \hat{\mathbf{n}}_{ij})\right] \hat{\mathbf{n}}_{ij} \tag{13}$$

where

$Y_n = 4 E_{eff} \sqrt{\bar{R} \delta_n}/3$, $c_n = 2 e_{eff} \sqrt{5 B_n m_{eff}/6}$, $e_{eff} = \ln(e)/\sqrt{\ln^2(e) + \pi^2}$, $B_n = 2 E_{eff} \sqrt{\bar{R} \delta_n}$.

where $\hat{\mathbf{n}}_{ij}$ is the unit vector pointing from particle i to particle j, $e$ is the coefficient of restitution of the particles, $\bar{R} = R_i R_j/(R_i + R_j)$ represents the geometric mean radius of the i and j particle, $\mathbf{v}_{ij}$ represents the velocity of the particle i relative to velocity of the particle j, $E_{eff} = [(1-\sigma_1^2)/E_1 + (1-\sigma_2^2)/E_2]^{-1}$ is the effective Elastic modulus computed by each particle's Elastic modulus and Poisson ratio, $\delta_n = R_i + R_j - |R_{ij}|$ is the displacement in normal direction and $m_{eff} = m_im_j/m_i+m_j$ is the effective masses of the particles.

The contact force in tangential direction is given as [23]
$$\mathbf{F}^t_{ij} = -\mathbf{min}\left[\mu_s\left|\mathbf{F}^n_{ij}\right|, Y_t\left(\delta_t \cdot \hat{\mathbf{t}}_{ij}\right) - \mathbf{c}_t\left(\mathbf{v}_t \cdot \hat{\mathbf{t}}_{ij}\right)\right]\hat{\mathbf{t}}_{ij} \qquad (14)$$
where
$Y_t = 8S_{eff}\sqrt{\bar{R}\delta_n}$, $\mathbf{c}_t = 2e_{eff}\sqrt{5B_t m_{eff}/6}$, $B_t = 8S_{eff}\sqrt{\bar{R}\delta_n}$,
$S_{eff} = 1/\left[2(2-\sigma_1)(1+\sigma_1)/E_1 + 2(2-\sigma_2)(1+\sigma_2)/E_2\right]$

where $\delta_t$ is the tangential displacement vector given by $\delta_t = \int_{t_0}^{t} v_t dt$, $\hat{\mathbf{t}}_{ij}$ is the unit vector along the tangential direction, $t_0$ is the time when the two particles just contact without deformation, $t$ is the time of collision. The tangential relative velocity is given by $v_t = [(v_i - v_j) \cdot \hat{\mathbf{t}}_{ij}]\hat{\mathbf{t}}_{ij} + (\omega_i \times R_i - \omega_j \times R_j)$ where $\omega_i$ or $\omega_j$ is the angular velocities of particles i or j and $\mathbf{R}_i$ or $\mathbf{R}_j$ is the radial vector.

For the cohesive force Johnson-Kendall-Roberts (JKR) model [24] which is based on Hertz elastic theory is used. The mathematical expression is given as follow:
$$\mathbf{F}_{coh} = \gamma\pi(D-R_i-R_j)(D+R_i-R_j)(D-R_i+R_j)(D+R_i+R_j)/4D^2 \qquad (15)$$
where $\gamma$ is the surface energy density and $D$ is the central distance between the i and j particles.

The torque due to rolling friction and tangential force is given by [25]:
$$\mathbf{T}^t_{ij} = \mathbf{R}_i \times \mathbf{F}^t_{ij} \qquad (16)$$
$$\mathbf{T}^r_{ij} = \mu_r \bar{\mathbf{R}} Y_n \delta_n \frac{\omega_{ij} \cdot \hat{\mathbf{t}}_{ij}}{|\omega_{ij}|}\hat{\mathbf{t}}_{ij} \qquad (17)$$
where $\mu_r$ is the rolling friction coefficient and $\omega_{ij} = \omega_i - \omega_j$ is the relative angular velocity.

The energy equation for fluid phase is [26]:
$$\frac{\partial}{\partial t}(\rho c_p \alpha_f T_f) + \nabla \cdot (\rho c_p \alpha_f U T_f) = -\nabla \cdot (\alpha_f k_f \nabla T_f) + \dot{S}_q + \frac{h_s A_s}{V_c}(T_f - T_s) \qquad (18)$$
where $c_p$ is the mean specific heat capacity of the fluid, $k_f$ is the mean thermal conductivity of the fluid, $\dot{S}_q$ is the source term due to evaporation, $h_s$ is the mean convection heat transfer coefficient for particle, $A_s$ is the surface area of the particle, $V_c$ is the volume of the CFD cell and $T_f$ and $T_s$ is the temperature of fluid and particle respectively.

The convection heat transfer coefficient $h_s$ is calculated using the Nusselt number correlation for particles in a stream of fluid [26].
$$Nu_s = 2 + 0.6\alpha_f^n Re_s^{1/2} Pr^{1/3} \qquad Re_s < 200 \qquad (19)$$
$$Nu_s = 2 + 0.5\alpha_f^n Re_s^{1/2} Pr^{1/3} + 0.02\alpha_f^n Re_s^{0.8} Pr^{1/3} \qquad 200 < Re_s < 1500 \qquad (20)$$
$$Nu_s = 2 + 0.000045\alpha_f^n Re_s^{1.8} \qquad Re_s > 1500 \qquad (21)$$
where n = 3.5, $Nu_s = h_s d/k_f$, $Re_s = \rho d|\mathbf{U}-\mathbf{u}_p|/\mu$ and $Pr = \mu c_p/k_f$.

Heat transfer between particles is also taken into consideration [23].
$$\dot{Q}_{cond,i-j} = h_{c,i-j}\Delta T_{i-j} \qquad (22)$$

$$h_{c,i-j} = \frac{4k_{si}k_{sj}}{k_{si}+k_{sj}}\left(A_{contact,i-j}\right)^{1/2} \qquad (23)$$

where $h_{c,i-j}$ is heat transfer coefficient due to conduction, $\Delta T_{i-j}$ is the temperature difference between particle i and j, $k_{si}$ is the thermal conductivity of particle i and $A_{contact, i-j}$ is the contact area for particle i and j.

The particle temperature is finally calculated as follow:
$$m_s c_{ps}\frac{dT_{s,i}}{dt} = h_s A_s\left(T_f - T_s\right) + \sum_{contacts\ i-j}\dot{Q}_{cond,i-j} \qquad (24)$$

The heat flux at the phase boundary for deviation of interfacial temperature $T_i$ from $T_{sat}$ is given as [27],
$$q_{ev} = h_{ev}\left(T_i - T_{sat}\right) \qquad (25)$$
where $q_{ev}$ is the heat flux due to evaporation and $h_{ev}$ is the interfacial vaporization heat transfer coefficient. It is given by [27],
$$h_{ev} = \frac{2C_e}{2-C_e}h_e^2\sqrt{\frac{M}{2\pi R}}\frac{\rho_2}{T_{sat}^{3/2}} \qquad (26)$$

Table 1 Simulation parameter and properties

| Parameters | Values |
| --- | --- |
| CFD domain size, mm×mm×mm | 6×6×6 |
| CFD cell size, mm×mm×mm | 0.1×0.1×0.1 |
| CFD time step | $1\times10^{-5}$ s |
| DEM time step | $1\times10^{-8}$ s |
| Particle number | 4500 |
| Particle diameter, radius | $5\times10^{-6}$ m |
| Particle density, $\rho_s$ | $7.87\times10^3$ kg/m$^3$ |
| Water (liquid) density, $\rho_1$ | 958.4 kg/m$^3$ |
| Vapor density, $\rho_2$ | 0.586 kg/m$^3$ |
| Water (liquid) thermal conductivity, $k_{f1}$ | 0.691 W/m K |
| Vapor thermal conductivity, $k_{f2}$ | 0.0246 W/m K |
| Particle thermal conductivity, $k_s$ | 80 W/m K |
| Water (liquid) viscosity, $\mu_1$ | $2.82\times10^{-4}$ N s/m$^2$ |
| Vapor viscosity, $\mu_2$ | $1.27\times10^{-5}$ N s/m$^2$ |
| Molecular mass of water, M | $18.052\times10^{-3}$ kg |
| Universal gas constant, R | 8.314 J/mol K |
| Mean free path of water, $\lambda$ | $2.5\times10^{-10}$ m |
| Specific heat of water (liquid), $c_{p1}$ | 4219 J/kg K |
| Specific heat of vapor, $c_{p2}$ | 2060 J/kg K |
| Specific heat of particle, $c_{ps}$ | 450 J/kg K |
| Surface tension of water, $\sigma$ | 0.07 N/m |
| Evaporation coefficient, $C_e$ | 0.1 |
| Saturation temperature, $T_{sat}$ | 373.15 K |
| Young's modulus, E | $200\times10^9$ N/m$^2$ |
| Restitution coefficient, e | 0.75 |
| Sliding friction coefficient, $\mu_s$ | 0.42 |
| Rolling friction coefficient, $\mu_r$ | $2\times10^{-4}$ |
| Poison ratio, $\sigma$ | 0.29 |
| Surface energy density, $\gamma$ | 0.280 J/m$^2$ |

where $C_e$ is the evaporation coefficient, $h_e$ is the enthalpy of vaporization, $\rho_2$ is the density of vapor, $R$ is the universal gas constant and $M$ is the molecular mass of water. The gradient of the volume fraction field $\alpha_1$ is zero everywhere except at the interface. The interfacial area can be calculated by taking the volume integral of the magnitude of the $\alpha_1$ field over a region encompassing the interface [27]. This gives us a way to define the evaporation source term $\dot{S}_q$ and $\dot{S}_v$ based on equation (22). And the temperature source term due to evaporation is

$$\dot{S}_q = q_{ev} \int_V |\nabla \alpha_1| dV / V \quad (27)$$

The mass source term due to evaporation is

$$\dot{S}_v = \dot{S}_q / \rho_f h_e \quad (28)$$

## RESULTS AND DISCUSSION

In this study 4500 particles each with a radius of 5 μm are initially suspended in liquid water film without touching each other. The size of the CFD domain is 6 mm × 6 mm × 6 mm. Liquid water filled the container to a height of 0.5 mm. To prevent the influence of side walls, periodic boundary condition is applied in lateral directions. The time step for the DEM case is $1\times10^{-8}$ s whereas the time step for the CFD case is $1\times10^{-5}$ s which means that for each CFD iteration DEM runs for 1000 iterations. Smaller time step for DEM is applied to prevent any unrealistic overlap between particles during collisions. The entire CFD domain is kept at a temperature of 373.15 K initially with the exception of bottom wall which is kept at a temperature of 378.15 K for all the time.

Figure 1 shows the initial configuration of the fluids and particles.

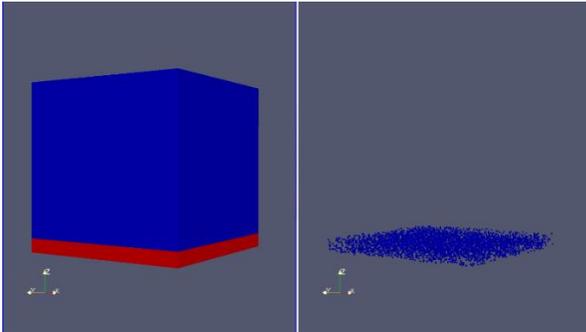

**Figure 1** Fluid and particle configuration at time t = 0 s

Figure 2 shows evaporation of water film from the surface. Liquid water is found to be completely evaporated at about 2.17 sec which means that evaporation still continues after the particles settle down. Particles are found to be completely settled at around 1 s which is much longer than the theoretical value to fall down if there is no water film. This is expected since there are drag force and buoyant force acting on the particles.

Figure 3 shows the packing structure of the particles after they settled down at the bottom of the container. It can be seen that packing configuration tends to form a pattern with the particles agglomerated in a special way. This clustering of particles continues as the water continues to evaporate.

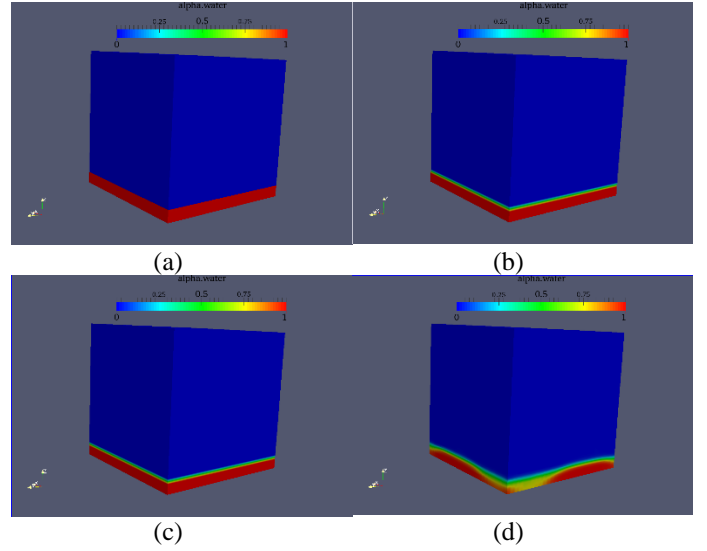

**Figure 2** Volume fraction alpha1 at (a) 0 s (b) 0.529 s (c) 1.39 s and (d) 2.04 s

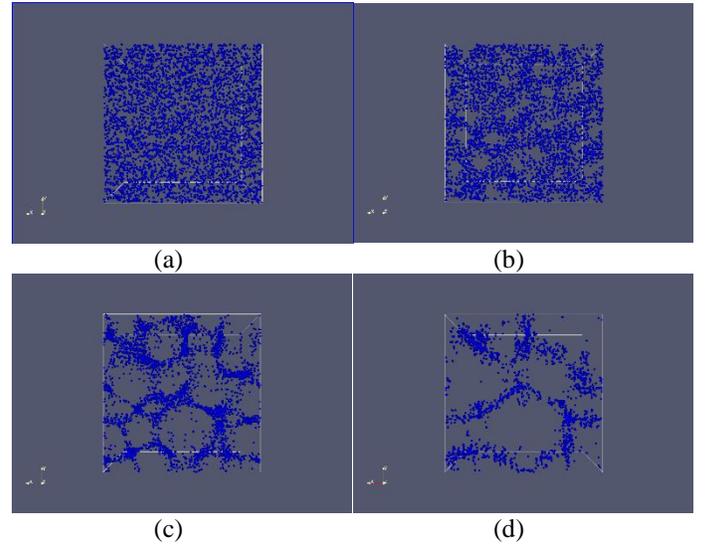

**Figure 3** Arrangement of particles at (a) 1 s (b) 1.9 s (c) 2.0 s and (d) 2.1 s

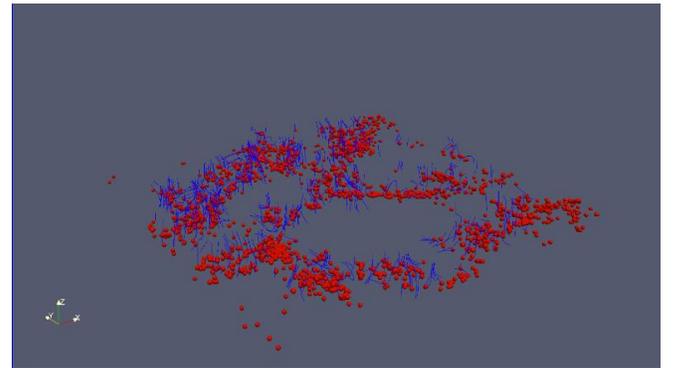

**Figure 4** Streamlines of resultant implicit forces on particles

Figure 4 shows the streamlines (blue lines) of the resultant forces acting on the particles at 2 s. It can be seen that lines are prominent around the particle clusters. To characterize the packing structure of the particles, the distribution of coordination number and radial distribution function (RDF) are taken into account. To define particles that are touching each other a critical distance equal to $1.01(r_1+r_2)$, where $r_1$ and $r_2$ are the radii of two particles, is set. Coordination numbers of the particles are calculated based on the assumption that if the center to center distance between two particles is less than this critical distance then the particles are considered as touching each other.

Figure 5 shows the distribution of coordination number at various times. At time equal to 1 s particles settled down at the bottom but they are barely touching each other. This can be easily seen from the Figure 5 (a) where most particles are shown to have a zero coordination number. At time equal to 2.05 s there is a gradual rise in coordination number which shows that some particles are touching and coming closer to each other. At times equal to 2.135 s and 2.167 s coordination numbers as high as 8 or 9 are observed. This means that some particles are sitting in a closed packed region.

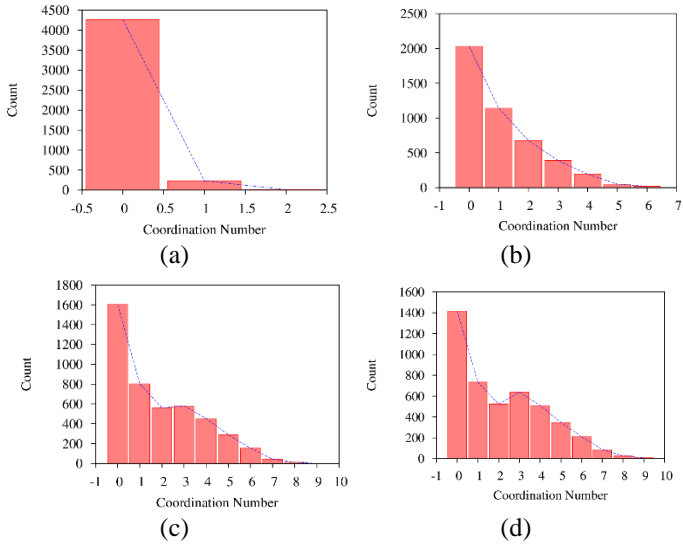

**Figure 5** Distribution of coordination number at (a) 1 s (b) 2.05 s (c) 2.135 s and (d) 2.167 s

Another parameter which is used to analyze the packing structure is the radial distribution function (RDF). It describes the probability to find a particle in a shell $dr$ at a distance r from another reference particle.

$$g(r) = \frac{V}{N} \frac{dN(r)}{4\pi r^2 dr} \quad (29)$$

where $V$ represents the total volume occupied by particles, $N$ is the total number of particles, and $dN(r)$ is the number of particles in a shell of width $dr$ at distance r from the reference particle.

RDF at time equal to 1 s shown in the Figure 6 (a) shows no peak since the particles are not touching each other. At time 2.05 s RDF shows peaks indicating the fact that particles are now forming clusters. A closer look at the peaks reveals that there are actually three peaks. The first peak is at $2r$ which corresponds to the one to one contact configuration. The second and third peak are at $2\sqrt{2}r$ and $4r$ which correspond to the edge-sharing-in-plane equilateral triangle and three particles centers in a line contact type. The peak values of RDF continue to change as the particle packing structure changes.

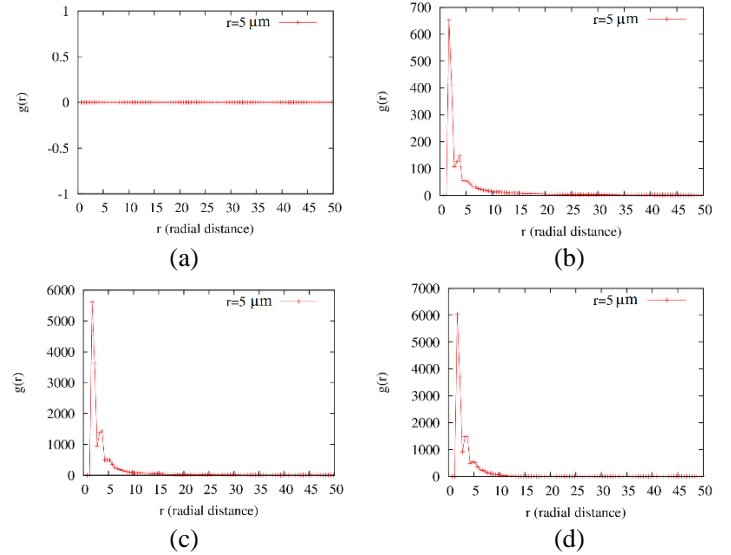

**Figure 6** Radial distribution function (RDF) at (a) 1 s (b) 2.05 s (c) 2.135 s and (d) 2.167 s

## CONCLUSION

A numerical investigation on evaporation induced self-assembly of sub-micron particles is carried out using a coupled CFD-DEM approach. The interaction between fluid and particle is thoroughly considered by taking into account the momentum exchange and heat transfer between particle and fluid. In the simulation liquid water film is allowed to evaporate and leave the particles at the container alone. Interesting patterns are seen to emerge as the liquid water film evaporates. The resulting packing structure is analyzed in terms of the range of coordination number and radial distribution function which also indicate the self-assembly of the particles.

## ACKNOWLEDGMENTS

Support for this work by the U.S. National Science Foundation under grant number CBET-1404482 is gratefully acknowledged.

## REFERENCES

[1] Oulton, R. F., Stavrinou P. N., Willams R., et al. Optical Microstructures Formed by Self-Assembly During Mbe Re-Growth of Gaas on Pre-Patterned Substrates. Quantum Electronics and Laser Science Conference, 2000 (QELS 2000) Technical Digest; 2000 12-12 May 2000; 2000. p. 55-56.

[2] Li, F., Wu M. C., Choquette K. D., Crawford M. H. Self-Assembled Microactuated Xyz Stages for Optical Scanning and Alignment. Solid State Sensors and Actuators, 1997 TRANSDUCERS '97 Chicago, 1997 International Conference on; 1997 16-19 Jun 1997; 1997. p. 319-322 vol.311.

[3] Roscoe, S. B., Yitzchaik S., Kakkar A. K., Marks T. J., Lin W., Wong G. K., "Ion Exchange Processes and Environmental Effects in Chromophoric Self-Assembled Superlattices. Manipulation of Microstructure and Large Enhancements in Nonlinear Optical Response," *Langmuir*, 1994; **10**(5): pp. 1337-1339.

[4] Feng, Y., Jiang H., Wang Y., et al., "Multi-Walled Carbon Nanotubes (Mwcnts)-Bridged Architecture of Ternary Bi2o3/Mwcnts/Cu Microstructure Composite with High Catalytic Performance Via two-Step Self-Assembly," *Solid State Sciences*, 2012; **14**(8): pp. 1045-1049.

[5] Deng, Y., Cai Y., Sun Z., et al., "Multifunctional Mesoporous Composite Microspheres with Well-Designed Nanostructure: A Highly Integrated Catalyst System," *Journal of the American Chemical Society*, 2010; **132**(24): pp. 8466-8473.

[6] Choi, S., Pisano A. P., Zohdi T. I., "An Analysis of Evaporative Self-Assembly of Micro Particles in Printed Picoliter Suspension Droplets," *Thin Solid Films*, 2013; **537**: pp. 180-189.

[7] Smith, J. S., Yeh H. J. J. Method for Fabricating Self-Assembling Microstructures. Google Patents; 1998.

[8] Mahoney, L., Koodali R., "Versatility of Evaporation-Induced Self-Assembly (Eisa) Method for Preparation of Mesoporous Tio2 for Energy and Environmental Applications," *Materials*, 2014; **7**(4): p. 2697.

[9] Brinker, C. J., Lu Y., Sellinger A., Fan H., "Evaporation-Induced Self-Assembly: Nanostructures Made Easy," *Advanced Materials*, 1999; **11**(7): pp. 579-585.

[10] Shao, F., Ng T. W., Efthimiadis J., Somers A., Schwalb W., "Evaporative Micro-Particle Self Assembly Influenced by Capillary Evacuation," *Journal of Colloid and Interface Science*, 2012; **377**(1): pp. 421-429.

[11] Ohtsuki, S., Matsuoka T., "Numerical Simulation of Solid Particle Behaviors in Fluid Flow by Using a Numerical Method Coupling Technique," *International Journal of the JCRM*, 2009; **4**(2): pp. 61-67.

[12] Xiong, Q., Madadi-Kandjani E., Lorenzini G., "A Lbm–Dem Solver for Fast Discrete Particle Simulation of Particle–Fluid Flows," *Continuum Mech Thermodyn*, 2014; **26**(6): pp. 907-917.

[13] Guo, Y., Wu C. Y., Thornton C., "Modeling Gas-Particle Two-Phase Flows with Complex and Moving Boundaries Using Dem-Cfd with an Immersed Boundary Method," *AIChE Journal*, 2013; **59**(4): pp. 1075-1087.

[14] Liu, D., Bu C., Chen X., "Development and Test of Cfd–Dem Model for Complex Geometry: A Coupling Algorithm for Fluent and Dem," *Computers & Chemical Engineering*, 2013; **58**: pp. 260-268.

[15] Jing, L., Kwok C. Y., Leung Y. F., Sobral Y. D., "Extended Cfd–Dem for Free-Surface Flow with Multi-Size Granules," *International Journal for Numerical and Analytical Methods in Geomechanics*, 2015: pp. n/a-n/a.

[16] Guadarrama-Lara, R., Jia X., Fairweather M., "A Meso-Scale Model for Fluid-Microstructure Interactions," *Procedia Engineering*, 2015; **102**: pp. 1356-1365.

[17] "Numerical Investigation of Dynamic Effects for Sliding Drops on Wetting Defects," *Physical Review E*, 2015; **91**(2).

[18] Ounis, H., Ahmadi G., McLaughlin J. B., "Brownian Diffusion of Submicrometer Particles in the Viscous Sublayer," *Journal of Colloid and Interface Science*, 1991; **143**(1): pp. 266-277.

[19] Dou, X., Mao Y., Zhang Y., "Effects of Contact Force Model and Size Distribution on Microsized Granular Packing," *Journal of Manufacturing Science and Engineering*, 2014; **136**(2): pp. 021003-021003.

[20] Zhang, H. P., Makse H. A., "Jamming Transition in Emulsions and Granular Materials," *Physical Review E*, 2005; **72**(1): p. 011301.

[21] Silbert, L. E., Ertaş D., Grest G. S., Halsey T. C., Levine D., Plimpton S. J., "Granular Flow Down an Inclined Plane: Bagnold Scaling and Rheology," *Physical Review E*, 2001; **64**(5): p. 051302.

[22] Brilliantov, N. V., Spahn F., Hertzsch J.-M., Pöschel T., "Model for Collisions in Granular Gases," *Physical Review E*, 1996; **53**(5): pp. 5382-5392.

[23] Kloss, C., Goniva C., Hager A., Amberger S., Pirker S., "Models, Algorithms and Validation for Opensource Dem and Cfd-Dem," *Progress in Computational Fluid Dynamics, an International Journal*, 2012; **12**(2): pp. 140-152.

[24] Johnson, K. L., Kendall K., Roberts A. D., "Surface Energy and the Contact of Elastic Solids," *Proceedings of the Royal Society of London Series A, Mathematical and Physical Sciences*, 1971; **324**(1558): pp. 301-313.

[25] Zhou, Y. C., Wright B. D., Yang R. Y., Xu B. H., Yu A. B., "Rolling Friction in the Dynamic Simulation of Sandpile Formation," *Physica A-statistical Mechanics and Its Applications*, 1999; **269**(2): pp. 536-553.

[26] Li, J., Mason D. J., "A Computational Investigation of Transient Heat Transfer in Pneumatic Transport of Granular Particles," *Powder Technology*, 2000; **112**(3): pp. 273-282.

[27] Hardt, S., Wondra F., "Evaporation Model for Interfacial Flows Based on a Continuum-Field Representation of the Source Terms," *J Comput Phys*, 2008; **227**(11): pp. 5871-5895.
6